# Observation of TeV Gamma Rays from the Crab Nebula with Milagro Using a New Background Rejection Technique

R. Atkins,[1] W. Benbow,[2,a] D. Berley,[3] E. Blaufuss,[3] J. Bussons[3,b], D. G. Coyne,[2] R.S. Delay,[9] T. DeYoung,[2] B. L. Dingus,[7] D. E. Dorfan,[2] R. W. Ellsworth,[4] A. Falcone,[5,c] L. Fleysher,[6] R. Fleysher,[6] G. Gisler,[7] M. M. Gonzalez,[1] J. A. Goodman,[3] T. J. Haines,[7] E. Hays,[3] C. M. Hoffman,[7] L. A. Kelley,[2] J. McCullough,[2,d] J. E. McEnery,[1,e] R.S. Miller,[5] A. I. Mincer,[6] M. F. Morales,[2,f] P. Nemethy,[6] D. Noyes,[3] J. M. Ryan,[5] F. W. Samuelson,[7] M. Schneider,[2] B. Shen,[8] A. Shoup,[9] G. Sinnis,[7] A. J. Smith,[3] G. W. Sullivan,[3] O. T. Tumer,[8] K. Wang,[8,g] M. Wascko,[8,h] D. A. Williams,[2] S. Westerhoff,[2,i] M. Wilson,[1] X. Xu,[7] G. B. Yodh[9]

[1]University of Wisconsin, Madison, WI 53706
[2]University of California, Santa Cruz, CA 95064
[3]University of Maryland, College Park, MD 20742
[4]George Mason University, Fairfax, VA 22030
[5]University of New Hampshire, Durham, NH 03824-3525
[6]New York University, New York, NY 10003
[7]Los Alamos National Laboratory, Los Alamos, NM 87545
[8]University of California, Riverside, CA 92521
[9]University of California, Irvine, CA 92717
[a]Now at Max Planck Institute, Heidelberg, Germany
[b]Now at Universite de Montpellier II, Montpellier, France
[c]Now at Purdue University, West Lafayette, IN
[d]Now at Cabrillo College, Aptos, CA
[e]Now at Goddard Space Flight Center, Green Belt, MD
[f]Now at Massachusetts Institute of Technology, Cambridge, MA
[g]Now at Standford Linear Accelerator Center, Menlo Park, CA
[h]Now at Louisiana State University, Baton Rouge, LA
[i]Now at Columbia University, New York, NY

Abstract

The recent advances in TeV gamma-ray astronomy are largely the result of the ability to differentiate between extensive air showers generated by gamma rays and hadronic cosmic rays. Air Cherenkov telescopes have developed and perfected the "imaging" technique over the past several decades. However until now no background rejection method has been successfully used in an air shower array to detect a source of TeV gamma rays. We report on a method to differentiate hadronic air showers from electromagnetic air showers in the Milagro gamma ray observatory, based on the ability to detect the energetic particles in an extensive air shower. The technique is used to detect TeV emission from the Crab nebula. The flux from the Crab is estimated to be $2.68(\pm 0.42^{stat} \pm 1.4^{sys}) \times 10^{-7} (E/1TeV)^{-2.59}$ m$^{-2}$ s$^{-1}$ TeV$^{-1}$, where the spectral index is assumed to be as given by the HEGRA collaboration.

## 1. Introduction

Ground-based gamma-ray astronomy has its roots in the pioneering work of the 1950's (Galbraith and Jelley 1953, Nesterova and Chudakov 1955). However it was not until the late 1980's that the first convincing detection of a source of TeV gamma rays was made with a ground-based instrument. The innovation that led to this discovery was the development of the "imaging" technique in atmospheric Cherenkov telescopes (Hillas 1985, Weekes and Turver 1977). This technique can distinguish air showers induced by gamma rays from those induced by hadrons (protons and heavier nuclei). The imaging technique categorizes air showers by the shape and orientation of the Cherenkov light pool as observed in the image plane of an optical telescope (an air Cherenkov telescope). This technique was used by the Whipple experiment to detect TeV gamma-ray emission from the Crab nebula (Weekes et al. 1989). Since the initial discovery of the Crab at least 5 other sources of TeV gamma rays have been detected (Hoffman et al. 1999, Ong 1998).

Despite the success of imaging air Cherenkov telescopes, they have several limitations. Because they are optical instruments, they can only observe the sky on clear, dark (moonless) nights (the typical duty cycle of these instruments is between 5% and 10%), and they can only observe a small fraction of the sky at any one time (of order $4 \times 10^{-3}$ sr). In contrast, an instrument that detects the air shower particles that reach the ground, known as an extensive air shower (EAS) array, can operate 24 hours a day and can simultaneously view the entire overhead sky. Past efforts to distinguish hadronic and gamma ray induced air showers in EAS arrays have relied on the identification of muons. The CASA and CYGNUS arrays used shielded detectors to identify muons present in hadronic air showers at energies above 100 TeV (Alexandreas et al. 1992, Borione et al. 1994). While the CASA array achieved very high levels of background rejection (rejecting 94% of the cosmic-ray background above 115 TeV, and 99.93% of the background above 1175 TeV, while retaining over 72% of the gamma ray showers), no signals were observed in their data (Borione et al. 1997). It is generally believed that the absence of observed sources at these high energies is due to the absorption of high-energy photons by the low-energy background radiation fields (3 K and infrared) and/or the steeply falling spectra of astrophysical sources.

The Milagro detector is an EAS array that is sensitive to much lower energy primary photons (>100 GeV, with a median energy near 4 TeV) and can therefore detect more distant sources (up to redshift ~0.3) and sources that have intrinsic upper limits to the energy of the gamma rays produced. Here we report on the development of a technique to identify and reject hadronic background events in Milagro. We demonstrate the efficacy of the technique with a detection of TeV emission from the Crab nebula.

## 2. The Milagro Detector

The Milagro gamma-ray observatory has 723 photomultiplier tubes (PMTs) submerged in a 24-million liter water reservoir. The detector is located at the Fenton Hill site of Los Alamos National Laboratory, about 35 miles west of Los Alamos, NM, at an altitude of 2630 m asl (750 g/cm$^2$). The reservoir measures 80m x 60m x 8m (depth) and is covered by a light-tight barrier. Each PMT is secured by a Kevlar string to a grid of sand-filled PVC sitting on the bottom of the reservoir. The PMTs are arranged in two layers, each on a 2.8m x 2.8m grid. The top layer of 450 PMTs (under 1.4 meters of water) is used primarily to reconstruct the direction of the air shower. By measuring the relative arrival time of the air shower across the array the direction of the primary cosmic ray can be reconstructed with an accuracy of roughly 0.75°. The bottom layer of 273 PMTs (under 6 meters of water) is used primarily to discriminate between gamma ray initiated air showers and hadronic air showers. For a more detailed description of Milagro see Atkins et al. (2001).

An EAS at ground level is composed primarily of electrons, positrons, and low energy gamma rays. The gamma rays outnumber the electrons and positrons in the air shower by a factor of ~4. Since a radiation length in water is ~36 cm, the bulk of the gamma rays convert to electrons and positrons before reaching the top layer of PMTs. The relativistic charged particles emit Cherenkov radiation in a 41° cone. The top layer of the array detects the Cherenkov radiation from the charged particles in the air shower with high efficiency (~50% of all electromagnetic particles that enter the pond are detected). The combination of high particle detection efficiency and a moderately high altitude gives Milagro a lower energy threshold than other air shower arrays. Milagro is sensitive to gamma rays with energies above 100 GeV. With a nominal trigger threshold of 60 PMTs (of the 450 in the top layer) the event rate is 1700 Hz.

The amount of light detected at a PMT depends upon the detailed trajectories of the particles in a region around that PMT. Since electromagnetic particles may penetrate close to the PMTs in the top layer of the pond, the fluctuations in the amount of detected light are large. Thus the top layer is not very useful for imaging the energy flow in an EAS. In contrast, only muons, showering hadronic particles and energetic electromagnetic particles (such as those near the core of an air shower) can penetrate to the bottom layer of the pond. Thus the bottom layer can be used to image the energy deposited by the air shower. The image of the EAS in the bottom layer of Milagro is a good discriminator against hadronic cosmic rays. The method described below, based on the "compactness" of the EAS in the bottom layer eliminates the bulk of the events initiated by the hadronic cosmic rays (~90%), while retaining ~50% of the events initiated by gamma rays.

## 3. Identification and Rejection of Hadronic Events

When a hadronic cosmic ray enters the atmosphere it interacts with the nuclei in the air. These interactions lead to the production of charged pions which can decay into muons and neutrinos. In addition, multi-GeV hadronic particles may also reach the ground. In contrast, when a gamma ray enters the atmosphere the interactions with the nuclei in the air are almost purely electromagnetic, resulting in an air shower that contains mostly lower energy electrons, positrons, and gamma rays.

Muons have great penetrating power; muons with energy above 1.2 GeV (at ground level) reach the bottom layer of PMTs. The muons that reach the bottom layer illuminate a relatively small number of neighboring PMTs. Hadrons shower in the detector also yielding large pulses in a small group of neighboring PMTs in the bottom layer. Monte Carlo simulations indicate that 80% of cosmic ray induced air showers and 6% of gamma ray induced air showers that trigger Milagro have at least one muon or hadron entering the pond.

Figure 1 shows six representative Monte Carlo events imaged in the bottom layer of Milagro. The simulations were generated over a zenith angle range of 0-45 degrees. (Throughout this paper an identical event selection is made on the Monte Carlo events as is made on the data: the number of struck PMTs in the top layer must be greater than 60, the number of PMTs used in the angular fit must be greater than 20, and for gamma rays, the events must be reconstructed within 1.2 degrees of their true direction.) CORSIKA 6.003 (Knapp and Heck 1993) is used to generate the extensive air shower and GEANT 3.21 is used to track the interactions of the particles through the water and to simulate the response of the detector. The simulated cosmic ray events are comprised of protons and helium nuclei: roughly 20% of the event triggers are due to helium nuclei. The top three events are gamma ray induced events and the bottom three are proton-induced events. The area of each square is proportional to the number of photoelectrons (PEs) registered in the corresponding PMT and the area is saturated at 300 PEs. The events were selected so that the proton events have similar properties (core location and number of PMTs struck in the top layer) to the gamma ray events directly above them. Table 1 gives the relevant parameters for each event. The final parameter *C* (for compactness) is defined in the next paragraph.

**Table 1 Properties of events shown in Figure 1. The events were selected to have similar core locations and a similar number of PMTs struck in the top layer.**

| Panel | Energy (TeV) | Core Position | $N_{Top}$ | $C$ |
|---|---|---|---|---|
| 1.a (gamma) | 1.84 | 1.5 m from pond | 326 | 3.2 |
| 1.b (gamma) | 0.89 | In pond | 115 | 4.55 |
| 1.c (gamma) | 17.0 | 60 m from pond | 292 | 6.0 |
| 1.d (proton) | 36.0 | 4 m from pond | 366 | 0.36 |
| 1.e (proton) | 0.13 | In pond | 180 | 0.03 |
| 1.f (proton) | 14.0 | 55 m from pond | 302 | 0.96 |

It can be seen that the gamma ray events have relatively smooth PE distributions in the bottom layer and the hadronic events have well-defined clumps of high-intensity regions. Since there is no optical barrier between the two layers of PMTs, light generated in the top layer can be registered in the bottom layer PMTs. To minimize the impact of this optical crosstalk between the top and bottom layers of PMTs a PE threshold is applied to the PMTs in the bottom layer. Four different threshold values have been investigated: 1, 2, 3, and 4 PE thresholds have been applied to the parameter $C=NB/PEMax_B$. *NB* is the number of PMTs in the bottom layer with a pulse height above the PE threshold and $PEMax_B$ is the number of PEs in the PMT with the maximum number of PEs in the bottom layer. This parameter should yield a small value for events where a few PMTs in the bottom layer are struck with a large amount of light and a high value for events where many PMTs are struck with a small amount of light. For each threshold one may determine a *C* value below which events are characterized as hadron-like and rejected. The value of such a cut is determined by calculating the ratio of the significance of a signal after the cut is applied to the significance of the signal before the cut is applied to the data. This ratio is commonly referred to as the quality factor (*Q*) of the cut. For large numbers of events $Q=\varepsilon_\gamma/\sqrt{\varepsilon_p}$, where $\varepsilon_\gamma$ is the fraction of gamma ray events retained and $\varepsilon_p$ is the fraction of hadron induced events retained.

Because the quality factor is poor for a PE threshold of 1, and the fraction of gamma ray events retained is 20% or less for PE thresholds of 3 and 4, a PE threshold of 2 is used for the data analysis and in the remainder of this paper. The resulting parameter, *C*, is referred to as *compactness* in the remainder of the paper and in Table 1.

Figure 2 shows the compactness distribution for gamma ray events and cosmic ray events (from Monte Carlo simulations) along with the observed distribution from the data. There is reasonable agreement between the Monte Carlo cosmic-ray events and the data. The *C* distribution for the data has slightly fewer events at larger values of C than the simulations predict and a larger fraction of the Monte Carlo events have C values near zero. The precise shape of the *C* distribution for the data is sensitive to the absolute value assigned to *PEMax$_B$* and therefore to the absolute calibration of the detector. Such an absolute calibration is difficult and in practice it is observed that the *C* distribution of the data changes slightly with detector calibrations, detailed trigger conditions, and the number of non-working PMTs. These variations have an insignificant impact upon the overall quality factor of the cut, which is sensitive only to the fraction of events reconstructed with a *C* value of 2.5 (see below). Over the observed detector variations this fraction changes by at most 2%.

Figure 3 shows the efficiencies for retaining data, cosmic rays, and gamma rays as a function of *C* and Figure 4 shows the *Q* value as a function of the minimum value of *C* required to retain an event. In each of these distributions there is generally good agreement between the Monte Carlo cosmic-ray events and the data. Requiring events to have C>2.5 rejects 90% of the simulated cosmic ray-induced air showers that trigger Milagro, 91.5% of the data (for this data sample), and retains 51% of the gamma ray induced air showers. This results in a predicted Q value of 1.6 comparing Monte Carlo cosmic ray events to Monte Carlo gamma ray events, and 1.7 comparing the data to Monte Carlo gamma ray events.

## 4. Properties of Compactness

The efficiency of the compactness parameter is a function of the energy of the primary gamma ray. At low gamma-ray energies only a few PMTs in the bottom layer have more than 2 PEs. So events with a low value of *PEMax$_B$* may have *C*<2.5. In addition, at lower energies the only gamma rays that trigger the detector tend to have their cores on the pond. The core of an electromagnetic shower typically contains the most energetic particles in the air shower. These particles can penetrate close to the bottom layer and again lead to events with *C*<2.5. In contrast it is found that the *C* distribution for cosmic-ray induced events is relatively independent of energy. Figure 5 shows the efficiency of the compactness cut as a function of primary energy for gamma ray and cosmic ray initiated events. One sees that the cut does not reach 50% efficiency until ~4 TeV, close to the median energy for gamma-ray showers. The efficiency for cosmic-ray initiated events is nearly independent of energy. The same figure shows the triggered energy distribution of gamma rays generated on an $E^{-2.4}$ spectrum that are reconstructed within 1.2 degrees of their true direction.

The compactness is also a function of the position of the core of the air shower. In particular, events with cores that land on the pond tend to have smaller values of compactness (i.e. more hadron like) than events with cores off the pond. As a result the quality factor for events with cores on the pond is substantially smaller than for events with cores off the pond. Figure 6 shows the compactness distributions for protons and gamma rays (from Monte Carlo) for events with cores on the pond (top) and events with cores off the pond (bottom) and Figure 7 shows the quality factor as a function of the cut imposed on the compactness parameter (all events with compactness less than the x-axis value are rejected) for events with cores on and off the pond. The quality factor for events with cores off the pond is ~50% higher than for the events with cores on the pond. If one rejects all events with C<2.5, 35% of gamma ray events with cores on the pond survive while 63% of gamma ray events with cores off the pond survive. Similarly for cosmic rays, 9% of events with cores on the pond survive and 12% of events with cores off the pond survive. At present the detector is too small to reliably locate the cores of air showers, though an array of water tanks that surrounds the pond is currently under construction. This array will contain the bulk of the air showers that trigger Milagro and will be able to reliably locate the core of the EAS that trigger Milagro.

Finally, the dependence of compactness on the zenith angle of the air shower has been investigated. Since the energy threshold of the instrument is a function of zenith angle and the number of muons generated in a hadronic cascade rises with primary energy, one might expect the rejection to improve with larger zenith angles. Figure 8 shows the compactness distributions for four different ranges of zenith angles. Figure 8(a) shows the compactness for zenith angles less than 15 degrees, 8(b) for events with zenith angles between 15 and 25 degrees, 8(c) for events with zenith angles between 25 and 35 degrees, and figure 8(d) for events with zenith angles between 35 and 45 degrees. The distributions look relatively similar. The quality factor as a function of the compactness cut is shown in Figure 9 (a-d) for the four zenith angle ranges. The smaller three zenith angle ranges give very similar quality factors; only the bin at the largest zenith angles (35-45 degrees) gives a significantly different quality factor. However, at the compactness value selected for background rejection ($C>2.5$) all four zenith angle ranges give a similar $Q$. Table 2 gives the efficiencies (and quality factors) of each of the zenith angle ranges for cosmic ray and gamma ray events for $C>2.5$.

## 5. Application to the Crab Nebula

The background rejection method derived above is applied to a search for TeV gamma rays from the Crab nebula. While several different values for the flux have been reported all observations to date indicate that the flux from the Crab is constant. This feature makes the Crab a useful source for calibrating the sensitivity of different instruments.

The dataset begins on June 18, 1999 and ends on September 13, 2002. Because of detector down time the effective exposure during this time interval is 917 days. The hardware trigger in Milagro requires 60 PMTs to fire within 200 ns. The arrival time and pulse height are recorded for each hit PMT. The angular fit is an iterative one where PMTs that make a large contribution to the chi-squared of the fit are removed and the event is refit. The number of PMTs remaining in the fit is called *nfit*. The angular resolution of the detector is dependent upon *nfit*. (For details on the fitting procedure see Atkins et al. 2000.) For this analysis we reject events with *nfit*<20; retaining roughly 91% of the data. From studies of the space angle difference in the reconstructed direction from independent, interleaved (like a checkerboard) portions of the detector, the shape of the Moon shadow, and Monte Carlo simulations we estimate the angular resolution of the instrument to be 0.75 degrees for all events with *nfit*>20. For square bins the optimal bin is 2.8 times this wide (Alexandreas et al. 1993) or 2.1 degrees in declination ($\delta$) by 2.1/cos($\delta$) in right ascension.

The analysis proceeds by counting the number of events that fall within the right ascension (*ra*) and declination ($\delta$) bands defined by the bin centered on the Crab nebula. This number is than compared with an estimate of the background. To estimate the background a method known as direct integration is used. The event rate in the detector is a function of the local coordinates [hour angle (*ha*) and declination ($\delta$)] and time. The technique assumes that the acceptance of the detector (expressed in local coordinates) is independent of the trigger rate over a two-hour period and that the cosmic rays form an isotropic background. Then one may write

$$N_{\text{exp}}[ra,\delta] = \iint E(ha,\delta) R(t) \varepsilon(ha,ra,t)\, dt\, d\Omega,$$

where *E(ha,$\delta$)* is the efficiency or acceptance of the detector and is simply the probability that an event comes from the differential angular element d$\Omega$ = d(ha)d($\delta$). *R(t)* is the event rate of the detector as a function of time, and *$\varepsilon$(ha,ra,t)* is one if the hour angle, right ascension, and the sidereal time are such that the event falls within the right ascension, declination bin of interest and zero otherwise. The function $E(ha,\delta)$ is found by binning the local sky in 0.1 x 0.1 degree bins and counting (and normalizing) the number of events that fall within each bin (and pass any cuts made on the data) over the 2-hour integration period. The integral is evaluated by summing the efficiency map (after the appropriate rotation is applied to convert hour angle to right ascension) over the observed events that pass the cuts imposed on the data. The error in the estimate of the background arises from the error in the determination of the detector acceptance, $E(ha,\delta)$. The significance of the excess is calculated using the prescription of Li and Ma (1983), where the factor $\alpha$, the ratio of time on source to time spent off source, is 1/11.21 and we have excluded the region around the Crab from our estimation of the background. A period of 2 hours for the background was chosen to obtain sufficient statistics for the background estimation while minimizing the systematic errors arising from changes in the acceptance of the detector over the integration period.

The results of the Crab analysis are given in Table 3. The results are given both before and after the compactness cut is been applied to the data. Figure 10 shows a map of the region around the Crab nebula after the compactness cut is applied. The color scale corresponds to the number of standard deviations excess (or deficit) from each region of the sky. At each point in the figure the excess (or deficit) is integrated over a square bin of size 2.1 degrees in declination ($\delta$) and $2.1/\cos(\delta)$ degrees in right ascension. While the $C$ cut removed 88.5% of the data, in rough agreement with the Monte Carlo simulations, the efficiency for gamma rays appears to be greater than 1. This nonphysical result arises from the large fluctuations in the background level (the low significance of the observed excess in the data before the application of the compactness cut). These same background fluctuations also explain the apparently large (and unphysical) quality factor of 3.3. (Since the efficiency for background events to pass the compactness cut is well determined from the data, 11.5%, and the gamma ray efficiency is bounded by 0 and 1, the quality factor must be bounded by 0 and 2.9.) The low statistical significance of the excess in the uncut data precludes one from establishing firm limits on the gamma ray efficiency of the compactness cut and all that can be said is that the measurement is marginally consistent with the Monte Carlo prediction. The efficiency of the compactness parameter for gamma rays has also been investigated by examining the excess from the Crab as a function of compactness. For $C$ values above 1 there is good agreement between the data and the Monte Carlo.

Table 2 Efficiencies for cosmic ray and gamma ray events for the cut C>2.5 for the four ranges of zenith angles.

| Zenith angle range | Gamma-ray efficiency | Proton efficiency | Quality factor |
|---|---|---|---|
| 0-15 | 61% | 18% | 1.44 |
| 15-25 | 55% | 13% | 1.53 |
| 25-35 | 47% | 9% | 1.57 |
| 35-45 | 42% | 6% | 1.71 |

Table 3 Results of the analysis of data from the Crab nebula. The results are given for all data and for data that with C>2.5.

| Data Selection | ON Source | Background | Excess | Significance |
|---|---|---|---|---|
| All Data | 18,374,036 | 18,365,694 | 8342 | 1.9 $\sigma$ |
| C > 2.5 | 2,119,449 | 2,109,732 | 9717 | 6.4 $\sigma$ |

The observed excess can be used to estimate the flux of TeV gamma rays from the Crab nebula. The observed excess includes a small contamination from the source in the background estimation. This is a result of the finite angular resolution of the detector and the method used to estimate the background. Monte Carlo studies show that 16% of the signal events fall outside of the source bin but within the declination strip used to estimate the background. Therefore the background level should be lowered (or the excess should be increased) by 138 events (0.16 x 9717 $\alpha$). Since the energy resolution of Milagro is relatively poor it is not meaningful to fit the shape of the spectrum. Instead four different functions have been assumed for the spectral shape and the differential flux coefficient is determined for each of these spectral shapes. The four spectral functions are: dN/dE $\alpha$ $E^{-2.49}$ and dN/dE $\alpha$ $E^{-2.44-0.151\log_{10}(E)}$ from the Whipple collaboration (Hillas et al.1998), dN/dE $\alpha$ $E^{-2.59}$ from the HEGRA collaboration (Aharonian et al. 2000), and dN/dE $\alpha$ $E^{-2.62}$ from the Tibet collaboration (Amenomori et al. 1999). Since the response of Milagro is dependent upon zenith angle Monte Carlo simulations are used to estimate the effective area of Milagro as a function of energy, averaged over a transit of the Crab. For a source with spectrum of *f(E)* the double integral over energy and time

$$I_0 \iint A_\gamma(E, \theta(t)) f(E) \, dE \, dt \text{ events / day}$$

is evaluated for the three spectral functions *f(E)*. In the above integral *$A_\gamma(E,\theta)$* is the effective area for gamma rays with energy *E* arriving from a zenith angle $\theta$. The calculation of the effective area includes the effect of the background rejection criteria (*C*>2.5), the angular reconstruction (the reconstructed direction of the event must be within 1.2 degrees of the true direction and the number of PMTs used in the fit must be greater than 20), and the trigger requirement of the detector (number of PMTs in the top layer > 60). After determining the value of the integrand, the observed (corrected) excess of 10.7 events/day is used to determine *$I_0$*. The results for the four different spectral shapes are given in Table 4. The systematic errors arise from two effects: a possible miscalibration of the detector and a systematic error in the Monte Carlo simulation of the air shower and/or the response of the detector to the air shower. The first is straightforward to estimate. The use of old calibrations to reconstruct newer data results in a 10% change in the observed signal level. Systematic errors in the simulation are more difficult to determine and an exhaustive study has not been performed. The simulation of the air shower relies on knowledge of the particle interactions and the atmospheric model. The simulation of the response of the detector relies on knowledge of the optical properties of the water, the optical properties of the water/cover boundary, the sensitivity of the PMTs, and the response of the electronics. With the exception of the water cover interface these parameters have been measured and the water quality is routinely measured. In principle the displacement of the shadow of the Moon from its true position (due to the deflection of charged particles in the earth's magnetic field) can be used to calibrate the energy response of Milagro. This analysis will be the subject of a future paper. Here we take a somewhat conservative value of 20% as a possible systematic error in the energy scale determined by the Monte Carlo. This leads to a ~60% systematic error in the determination of *$I_0$* which dominates the systematic error of the instrument.

**Table 4** The measured flux from the Crab nebula assuming four different spectral shapes as determined by other instruments. The first error is statistical and the second error is systematic. In the case of the last column no systematic error is given in the reference.

| Energy Spectrum | $E^{-2.49}$ | $E^{-2.44-0.151\log10E}$ | $E^{-2.59}$ | $E^{-2.62}$ |
|---|---|---|---|---|
| $I_0 \times 10^{-7}$ (m$^{-2}$ s$^{-1}$ TeV$^{-1}$) (this data) | 2.38±0.38±1.2 | 2.71±0.44±1.4 | 2.68±0.42±1.4 | 2.65±0.42±1.4 |
| $I_0$ (from reference) | 3.20±0.17±0.6 | 3.25±0.14±0.6 | 2.79±0.022±0.5 | 8.2±1.6 |
| Reference | 1 | 1 | 2 | 3 |

References. – (1) Hillas et al. 1998; (2) Aharonian et al. 2000; (3) Amenomori et al. 1999

Table 4 indicates that while the flux measured by Milagro is in good agreement with the air Cherenkov experiments, there is a disagreement with the results from the Tibet air shower array. Including the systematic errors from Milagro the results disagree at the 2.6 σ level, the reference for the Tibet result (Amenomori et al. 1999) does not discuss systematic errors. Given the large systematic errors one can not make a definitive statement, but among the other three results the data seem to favor the softer spectra as measured by HEGRA or the curved spectrum as measured by Whipple.

## 6. Conclusions

The bottom layer of Milagro is an imaging calorimeter that can be used to measure the lateral distribution of energy deposited in Milagro. Hadronic cosmic rays generate air showers with penetrating particles that deposit localized clumps of energy in the Milagro detector. A simple algorithm to differentiate air showers induced by hadronic cosmic rays from those induced by gamma rays has been developed. This simple cut, based on a compactness parameter ($NB/PEMax_B$), improves the sensitivity of Milagro by a factor of ~1.7. We have used this cut to observe TeV gamma-ray emission from the Crab nebula. The measured flux is consistent with previous measurements by atmospheric Cherenkov telescopes, but it is inconsistent with the reported result from the Tibet air shower array.

This is the first demonstration of the ability of an EAS array to reject hadrons and enhance the significance of an observation of a source of TeV gamma rays. More complex techniques that utilize additional information to improve the background rejection capabilities of the instrument are under investigation. In addition, an outrigger array that is currently under construction will provide knowledge of the core location of each event, which is expected to further increase the discrimination power of Milagro. As Milagro is a new and unique type of instrument we are only beginning to understand its response to cosmic rays and gamma rays. As understanding of this new instrument improves further improvements to the sensitivity of Milagro are expected.

**Figure Captions**

Figure 1. Gamma ray (top 3 images) and proton (bottom 3 images) events imaged in the bottom layer of Milagro. The area of each square is proportional to the pulse height registered in each PMT. The proton and gamma ray events have been selected to have similar properties in the top layer of the detector. See the text and Table 1 for details.

Figure 2. The *C* distributions for Monte Carlo gamma ray showers (dashed line), Monte Carlo cosmic ray showers (dotted line), and data (solid line). All of the histograms have been normalized to have unity area.

Figure 3. The fraction of Monte Carlo gamma rays (dashed line), Monte Carlo cosmic rays (dotted line), and data (solid line) with C values greater than x-axis value.

Figure 4. The quality factor, *Q*, as a function of *C*. Events with *C* greater than x-axis value are retained. The dashed line compares Monte Carlo gamma rays to Monte Carlo cosmic rays and the solid line compares Monte Carlo gamma rays to data.

Figure 5. The solid (dotted) line shows the fraction of gamma rays (cosmic rays) retained by the compactness cut (C>2.5) as a function of primary energy. The dashed line shows the energy distribution of the gamma rays that trigger Milagro. The gamma rays were generated with an $E^{-2.4}$ spectrum.

Figure 6. The compactness distributions of protons (solid line) and gamma rays (dotted line). (a) is for events with their cores in the pond and (b) is for events with their cores off the pond.

Figure 7. The Q factor for events with their cores on the pond (lower curve) and off the pond (upper curve).

Figure 8. The compactness distributions for 4 zenith angle ranges of cosmic ray (solid line) and gamma ray (dotted line) showers. (a) is for zenith angles less than 15 degrees, (b) for zenith

angles between 15 and 25 degrees, (c) for zenith angles between 25 and 35 degrees, and (d) for zenith angles between 35 and 45 degrees.

Figure 9. Q factor as a function of the cut on compactness for 4 different zenith angle ranges. Bottom curve is for zenith angles less than 15 degrees, the solid squares for zenith angles between 15 and 25 degrees, the open circles for zenith angles between 25 and 35 degrees and the filled circles for zenith angles between 35 and 45 degrees.

Figure 10. A map of the statistical significances in the region around the Crab nebula. The scale on the right is in standard deviations. The larger circle at the center shows the integration region used around each point in the map. The small circle shows the location of the Crab nebula.

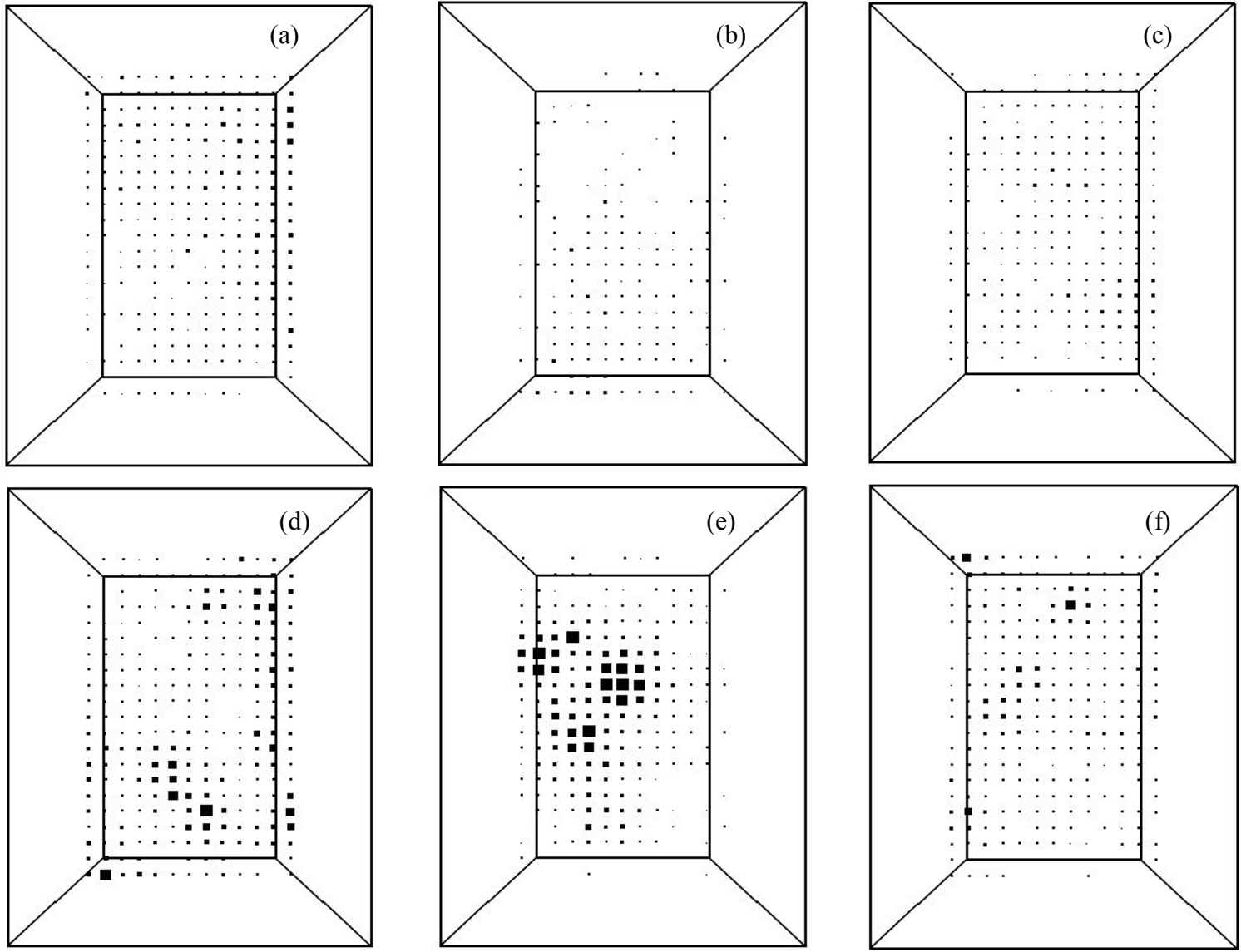

Figure 1

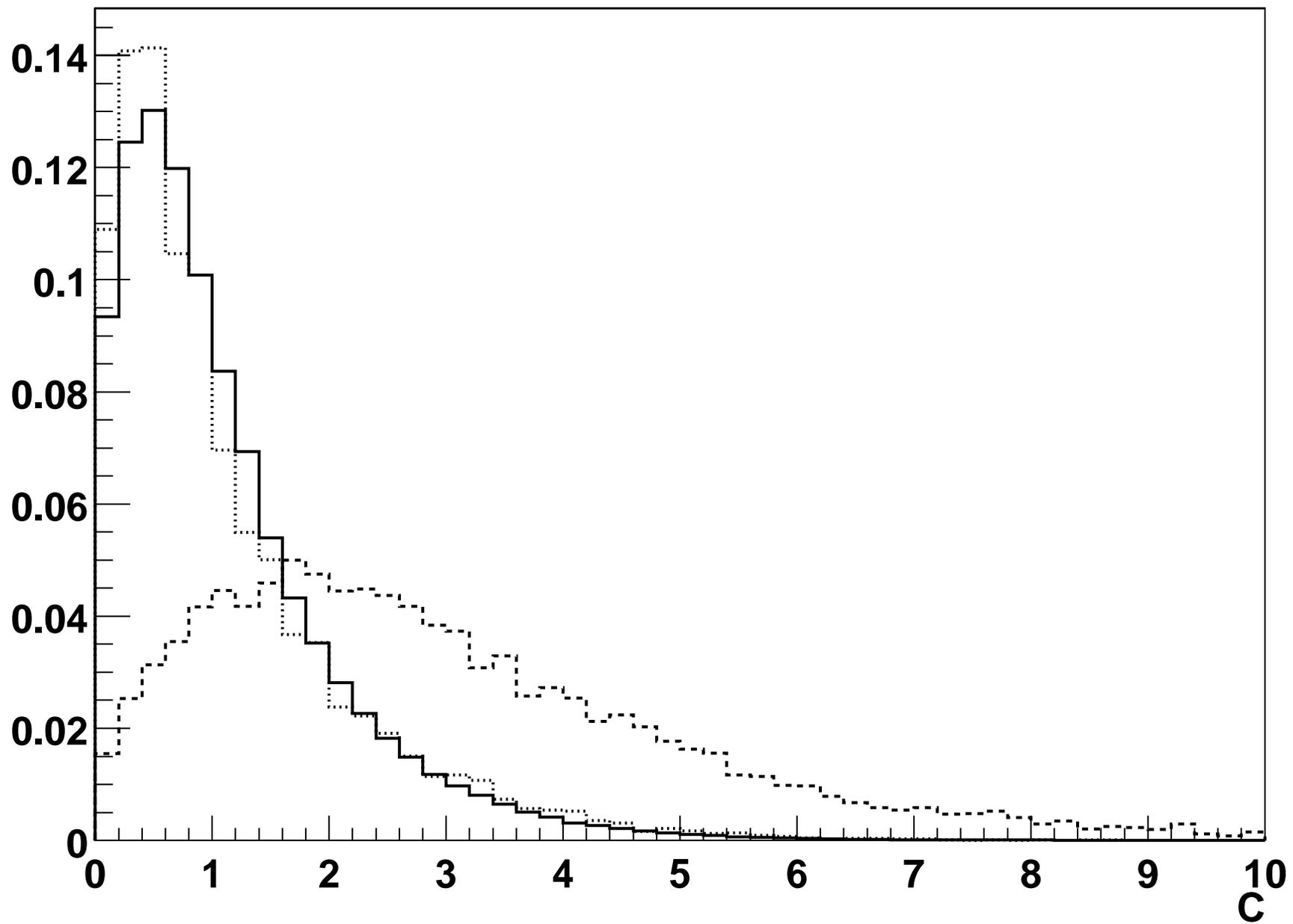

Figure 2

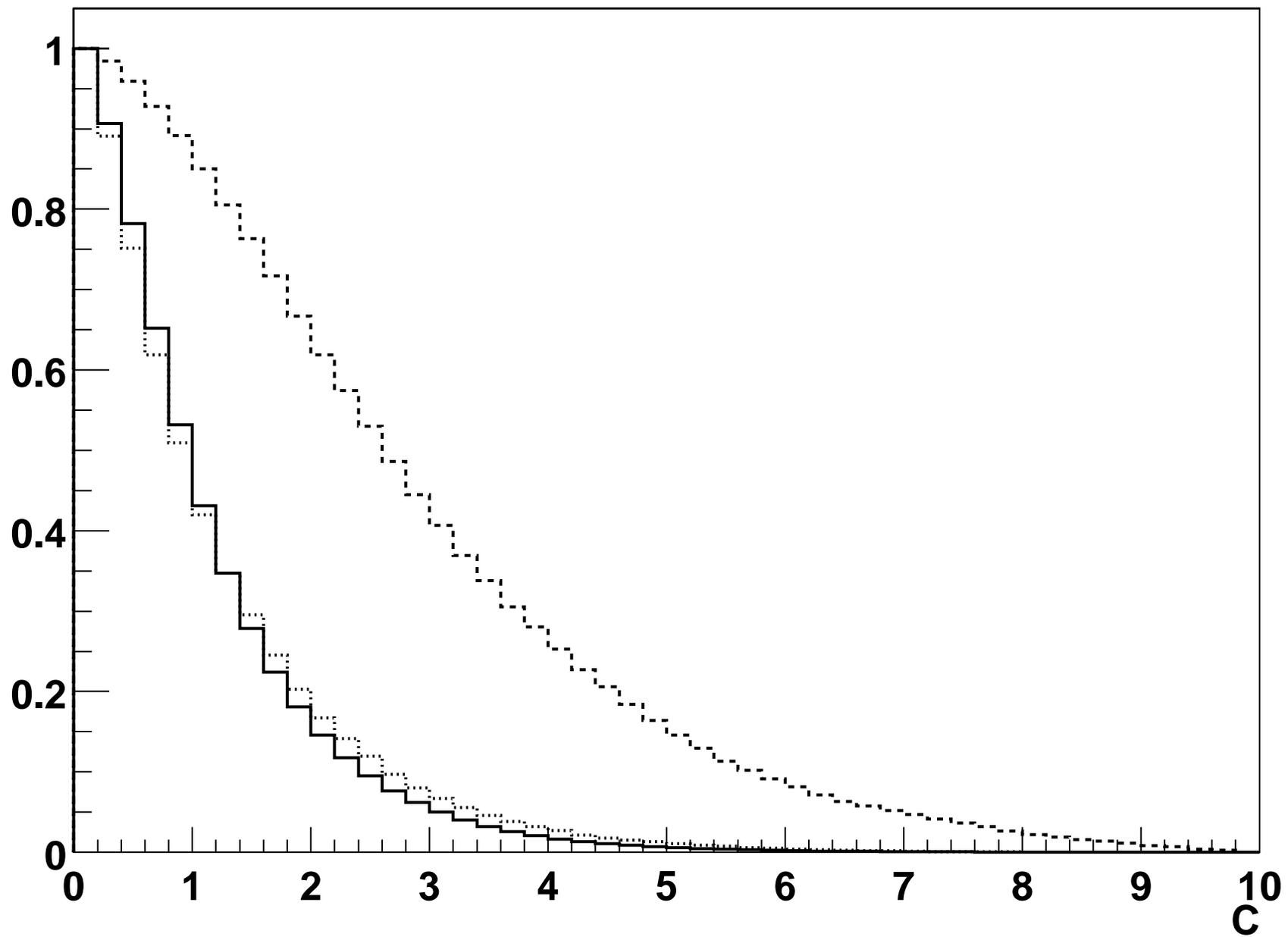

Figure 3

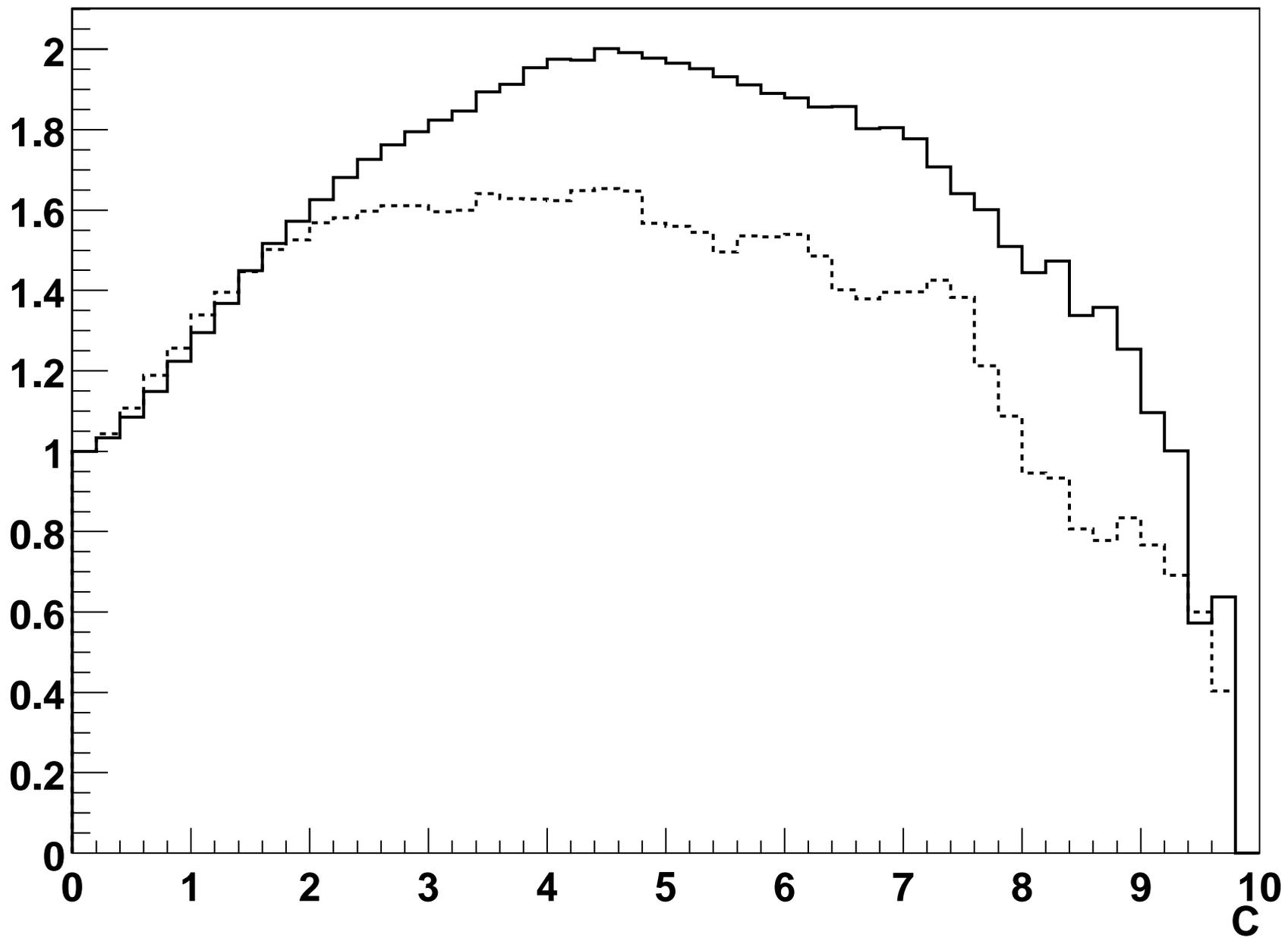

Figure 4

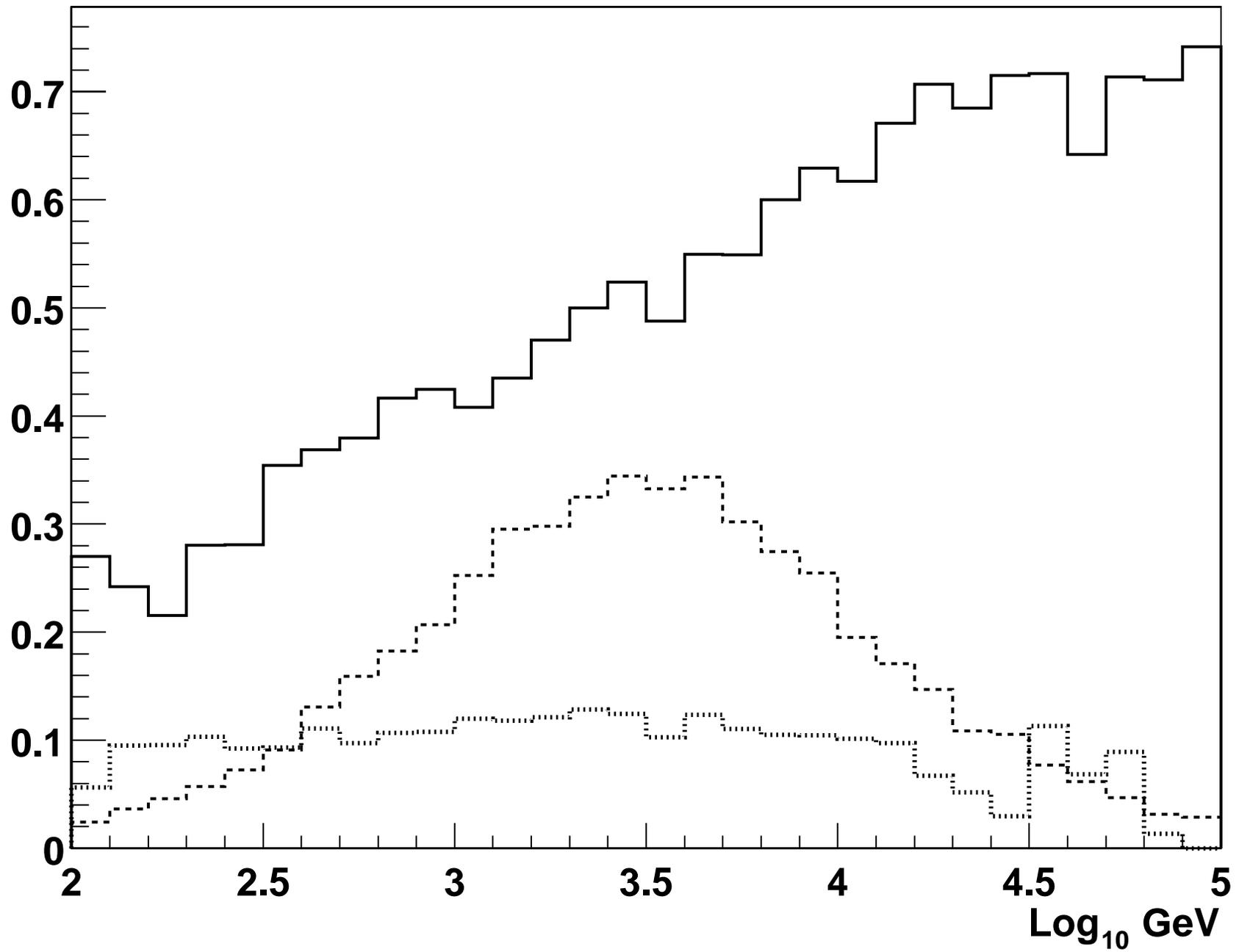

Figure 5

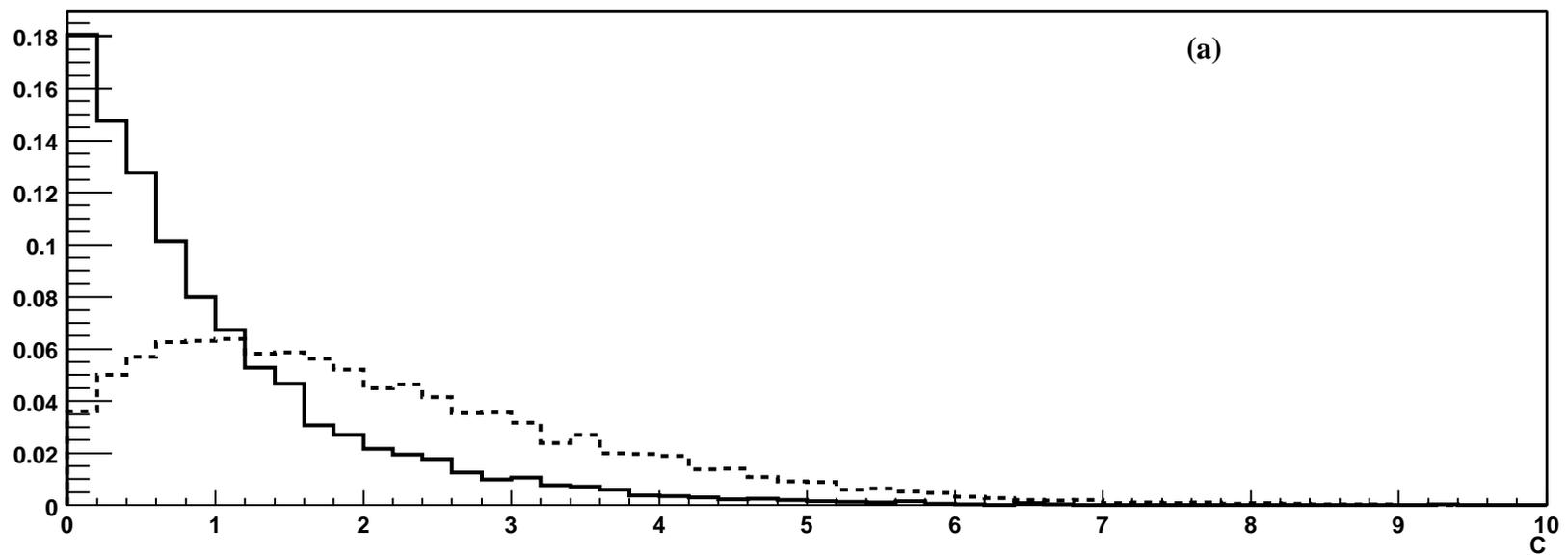

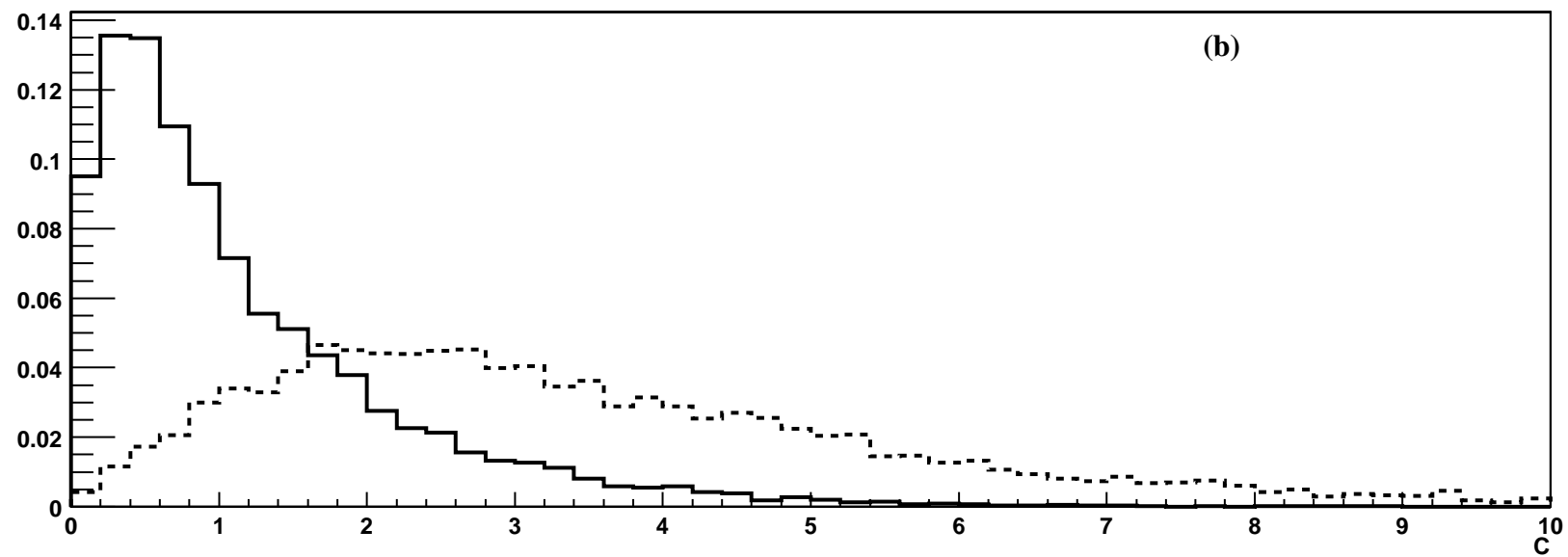

Figure 6

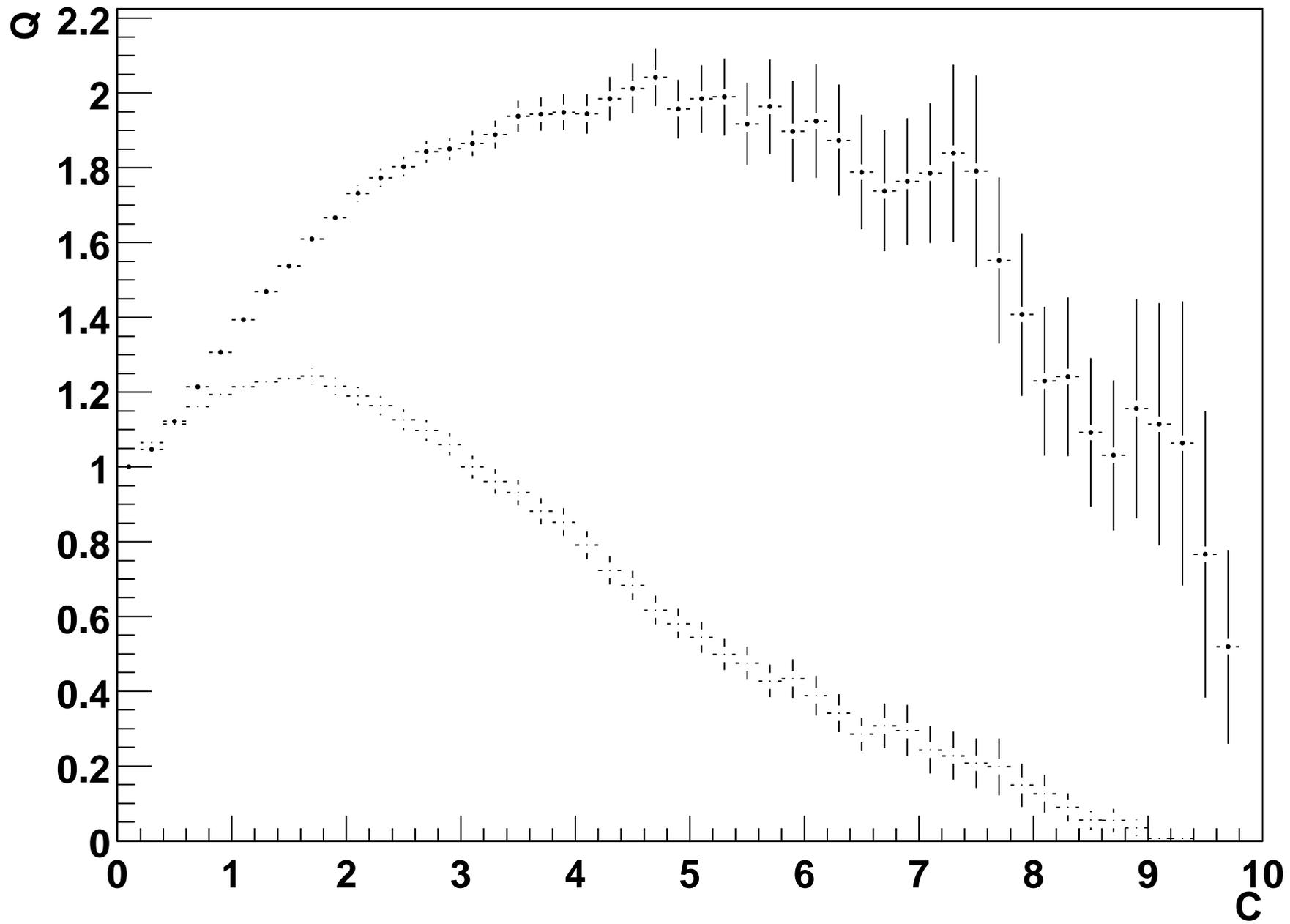

Figure 7

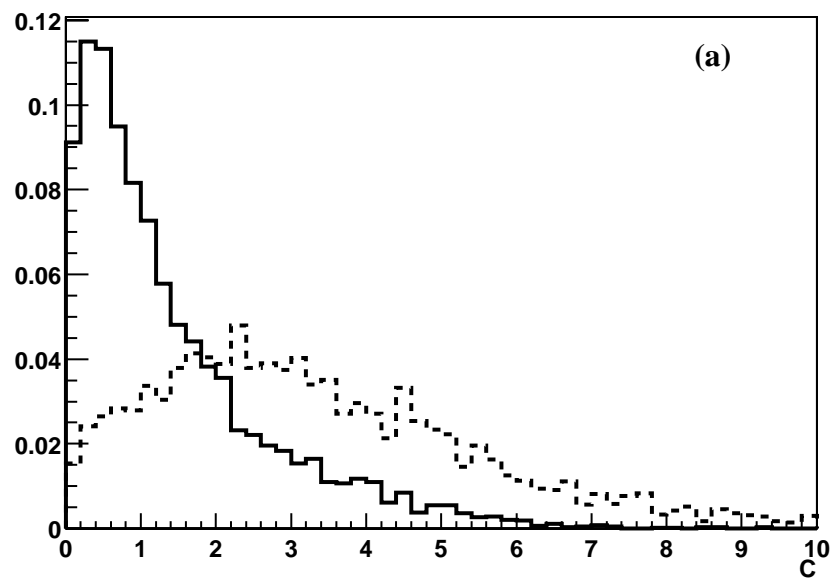
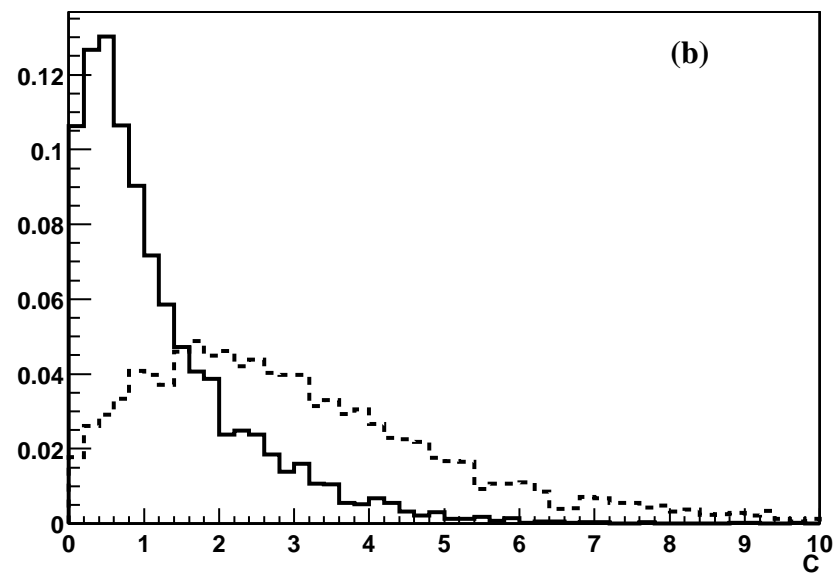
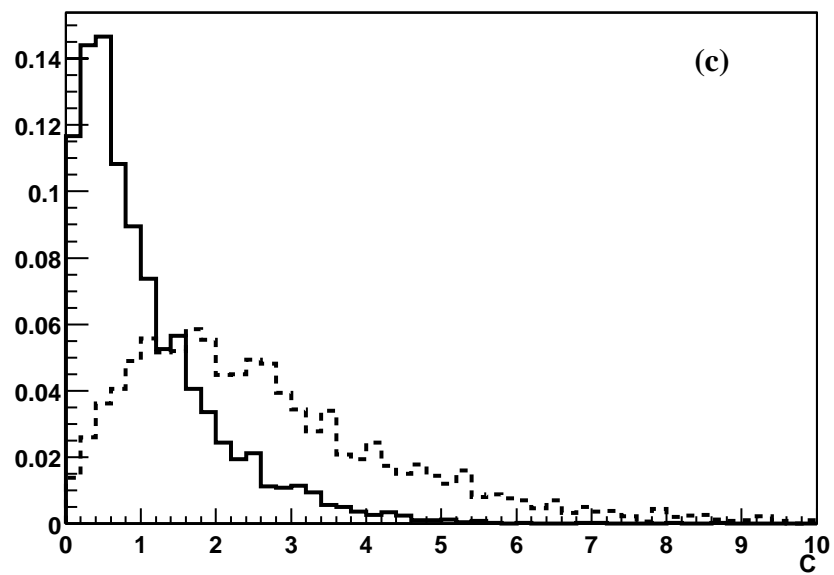
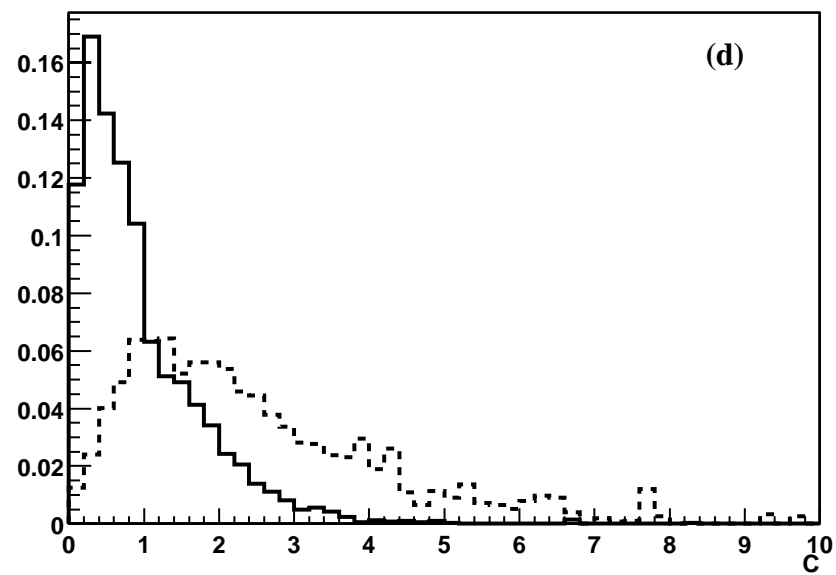

Figure 8

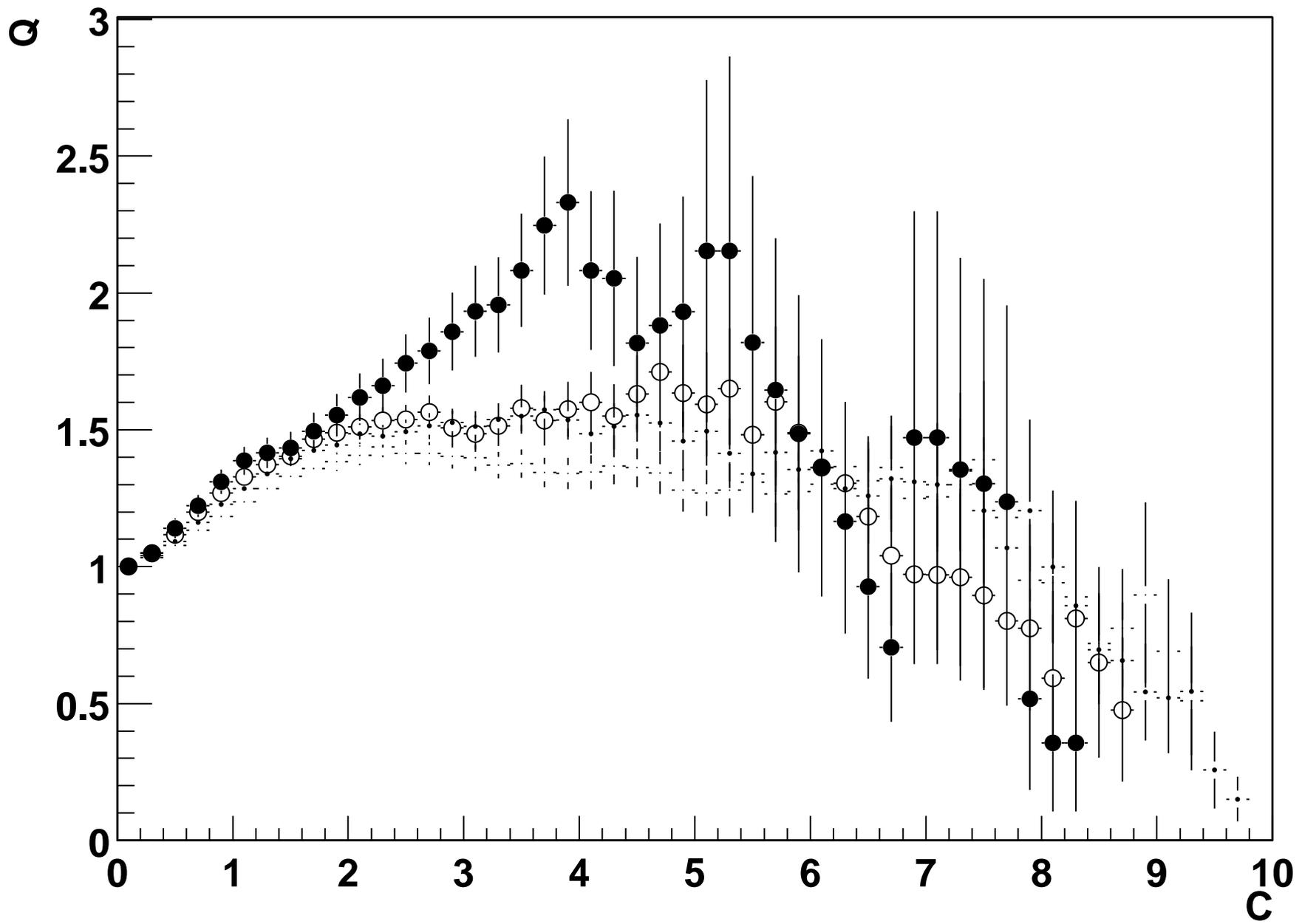

Figure 9

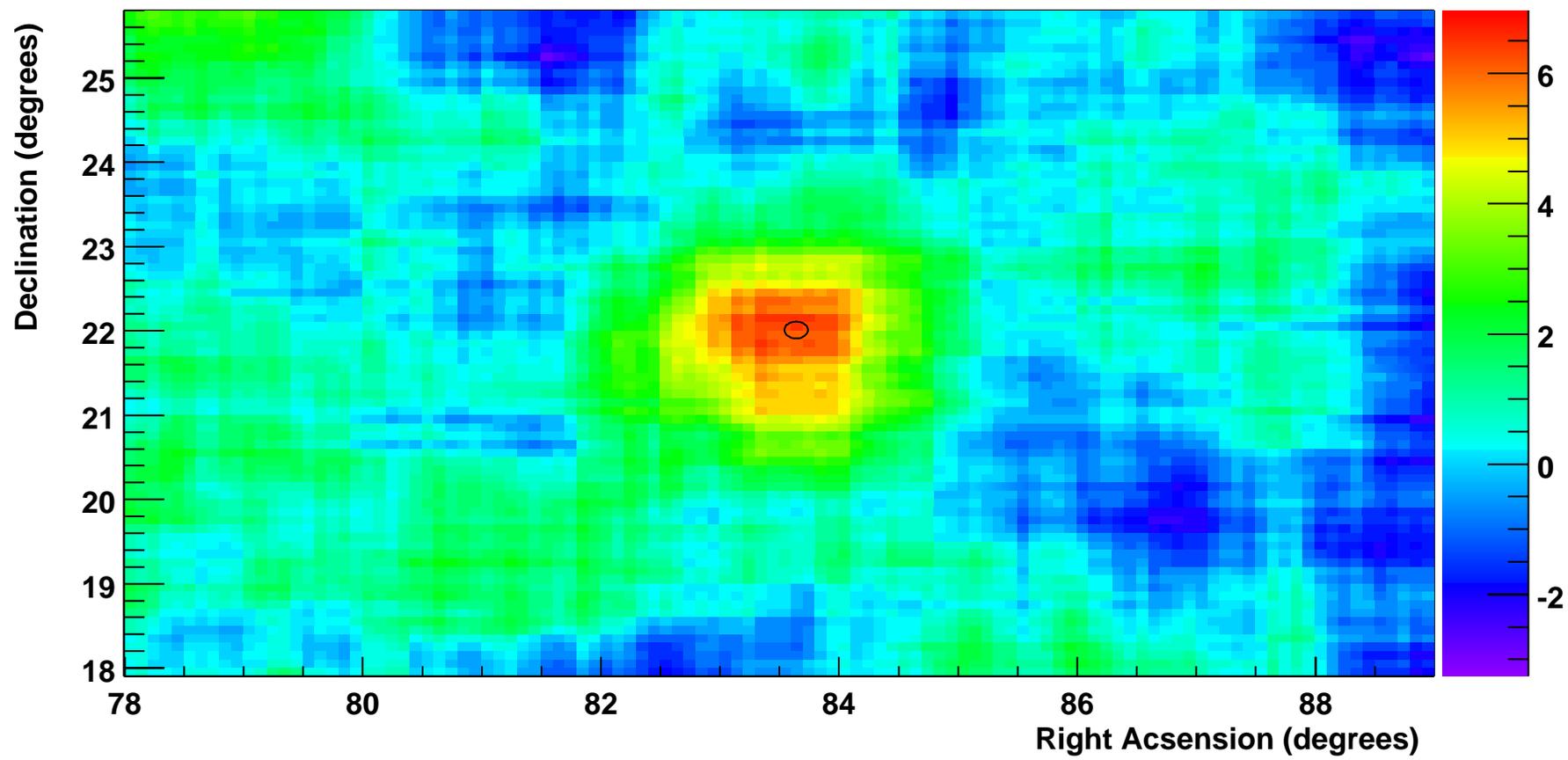

Figure 10 (color)